\documentclass[journal=jacsat,manuscript=article]{achemso}
\usepackage{geometry}
\usepackage{sidecap}
\usepackage{graphicx}% Include Figure files
\usepackage{caption}
\usepackage{dcolumn}% Align table columns on decimal point
\usepackage{bm}
\usepackage{mathtools}
\usepackage{amsmath}
\usepackage{chemmacros}
\usepackage{xcolor}
\usepackage{ulem}
\usepackage[version=4]{mhchem}
\usepackage[export]{adjustbox}
\newcommand{\EF}[1]{\textcolor{blue}{#1}}

\graphicspath{{/home/dino/Documents/Vesicle/ResubFigures/}}

\makeatletter
 \patchcmd{\acs@contact@details}{E}{*\,E}{}{}
\patchcmd{\acs@email@list@aux}{;}{\par*\,Email}{}{} % added <<<<<<<<<<<<
\makeatother

\author{Dino Osmanovi\'c}
\affiliation[UCLA]
{Department of Mechanical and Aerospace Engineering, University of California at Los Angeles, Los Angeles, CA, USA}
\email{osmanovic.dino@gmail.com}
\author{Elisa Franco}
\affiliation[UCLA]
{Department of Mechanical and Aerospace Engineering, University of California at Los Angeles, Los Angeles, CA, USA}
\alsoaffiliation[UCLABE]
{Department of Bioengineering, University of California at Los Angeles, Los Angeles, CA, USA}

\title{Generating forces in confinement via polymerization}

\begin{document}

\maketitle

\begin{abstract} 
Understanding how to produce forces using biomolecular building blocks is essential for the development of adaptive synthetic cells and living materials. Here we ask whether a dynamic polymer system can generate deformation forces in soft compartments by pure self-assembly, motivated by the fact that biological polymer networks like the cytoskeleton can exert forces, move objects, and deform membranes by simply growing, even in the absence of molecular motors. We address this question by investigating polymer force generation by varying the  release rate, the structure, and the interactions of self-assembling monomers.  First, we develop a toy computational model of polymerization in a soft elastic shell that reveals the emergence of spontaneous bundling which enhances shell deformation. We then extend our model to account more explicitly for monomer binding dynamics. We find that the rate at which monomers are released into the interior of the shell is a crucial parameter for achieving deformation through polymer growth. Finally, we demonstrate that the introduction of multivalent particles that can join polymers can either improve or impede polymer performance, depending on the amount and on the structure of the multivalent particles. Our results provide guidance for the experimental realization of polymer systems that can perform work at the nanoscale, for example through rationally designed self-assembling proteins or nucleic acids.
\end{abstract}

\section{Introduction}

Methods to build self-assembling molecules with arbitrary features have rapidly expanded in the past decade~\cite{krishna2024generalized}, prompting the development of new materials that may rival cells and tissues in their capacity for morphological adaptation~\cite{liu2022living}. For example, artificial DNA polymers can be activated dynamically through enzymatic reactions~\cite{green2019autonomous, agarwal2019enzyme}, organized via other DNA nanostructures~\cite{jorgenson2017self,agrawal2017terminating,zhang2023building}, and even functionalized with other compounds~\cite{mitchell2004self}. These polymers could be used to produce responsive materials that move and adapt by mimicking well-known biological mechanisms. Actin polymers, for instance, generate a variety of forces for directed motion~\cite{fletcher_introduction_2004} and for membrane deformation~\cite{simon_actin_2019} through mechanisms that include ratcheting~\cite{mogilner_force_2003,dominguez_actin_2011,rottner_actin_2017} and the coordination of polymers with proteins crosslinkers~\cite{lieleg_structure_2010}. The generation of forces by polymers in confinement has been studied through reconstituted cytoskeletal networks~\cite{limozin_organization_2003,tsai_chapter_2014, baldauf2022actomyosin,litschel2021reconstitution, bashirzadeh2021actin,sakamoto2024f} and through synthetic analogs of cytoskeletal polymers~\cite{kurokawa2017dna,agarwal2021dynamic,sauter2023artificial}. However, progress toward generating forces with synthetic systems is hampered by the scarcity of components to regulate polymerization, which play an important role in living cells. Further, actin polymerization is a nonequilibrium process that consumes ATP (actin monomers are ATP bound), and we lack artificial molecular motors comparable to myosin or kinesin. 
%Is it possible for a polymer system to produce forces even in the absence of complex processes for chemical energy conversion that have been evolved by cytoskeletal components?  

\begin{figure}[h]
    \begin{center}
    \includegraphics[width=0.8\textwidth]{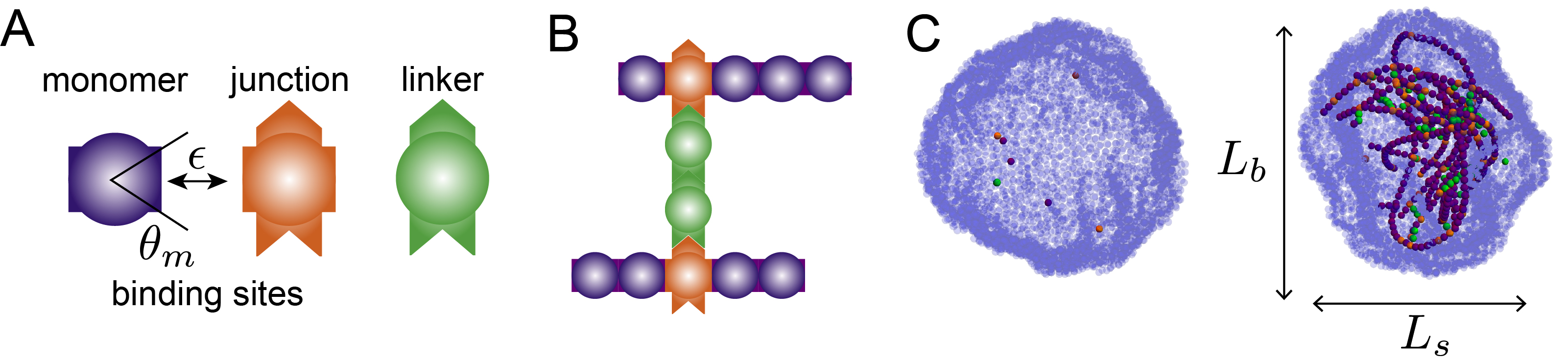}
    \caption{We seek to design the properties of monomers and their release rates such that they can generate a force against an elastic shell  A) We consider systems including  monomers with diverse valency and binding sites. Sites can bind with some energy $\epsilon$ and maximum angle $\theta_m$, as in equations \eqref{eq:fullpatch},\eqref{eq:squarewell},\eqref{eq:patch}, and  their shape represents whether they are compatible or not. Two blunt patches can bind together, as can the V-shaped patches, but blunt and V-shaped cannot interact.  B) Illustrative assembly resulting from the monomers in A). C) Our approach involves releasing different types of monomers into the interior of the elastic shell, leading to self assembly of structures that are dependent on both the structural elements and history of release. We ask what details of the assembly ``program'' can generate shell deformation.}
    \label{fig:fig1} %
    \end{center}
    \end{figure}

Given that we are mastering how to control the structure and the production of synthetic polymers, here we ask whether it may be possible to generate nonequilibrium forces by simply controlling the pool of structural elements available to polymerize at any given time. Are there general techniques by which polymerization of molecular components alone can generate controllable forces? Can we formulate particular ``molecular programs" defined by a pool of structural elements and the temporal instructions for their assembly, that can route a polymer network to produce forces of sufficient magnitude in a particular direction? We address these questions through computational modeling of monomer systems that form polymers in confinement. We consider {\it dynamic} polymerization with monomers of differing structure. Taking inspiration from theoretical studies of ``empty" fluctuating shells~\cite{paulose_fluctuating_2012,maji_network_2022}, we consider the problem of introducing axial deformation. We seek to identify molecular programs in which the monomer structure and the timing of monomer activation are sufficient control parameters to produce deformation forces purely via a passive self-assembly process. Our approach is summarized in Figure~\ref{fig:fig1}. 

Modeling work addressing related questions has been developed with the description of actin behavior in mind. For example, it has been demonstrated that crosslinking of actin can produce compressive forces~\cite{ma_crosslinking_2019}. Under confinement, it was shown that crosslinked actin networks tend to align along the longest axis of the compartment~\cite{akenuwa_organization_2023,adeli_koudehi_organization_2019}. Separate theoretical and computational studies under confinement have explored the behavior of semiflexible polymers~\cite{milchev_conformations_2017,nikoubashman_semiflexible_2017,fovsnarivc2013monte}, their mechanical properties~\cite{kierfeld_modelling_2010,claessens_microstructure_2006}, and  the possible deformations they can impart on soft compartments~\cite{shi2023morphological}. These studies demonstrate that interesting effects can occur when semiflexible polymers are confined within a radius comparable to their persistence length. However, these previous approaches usually consider polymers that have already reached a given length in confinement. To the best of our knowledge, previous work has not considered the role of polymerization in soft confinement, and have not examined the effects of varying monomer release rates as well as monomer interaction patterns.  

Our simulations demonstrate that polymerization in soft compartments can lead to axial deformation as long as monomer accumulation is slow and the assembled polymers are sufficiently stiff (high energetic penalty to polymer bending). We find that in these conditions the polymer system exhibits a transition from unbundled to bundled conformation. While bundled, the polymers become sufficiently strong to deform the shell, even if an individual polymer is not strong enough to do so. We also demonstrate that the introduction of multivalent  particles crosslinking the polymers can lead to compartment deformation, even for polymers which ordinarily wouldn't generate sufficient forces. 
%These findings suggequalitatively agrees with experimental work examining active assembly of cytoskeletal networks~\cite{sakamoto2024f}, confirming that how polymers interact is key to support the generation of forces. 

\section{Materials and Methods}
\subsection{Modelling Soft Compartments}

To study the generation of passive forces via self assembly of our monomers, we will simulate their interactions in a deformable elastic shell. For this purpose, we created a soft compartment by placing 2432 points (number of particles) uniformly on a sphere and having neighboring particles $i$ and $j$ interact via Hookean springs:
\begin{equation}
\phi_{V}(r_{ij})=\frac{1}{2}k_s (r_{ij}-d_{ij})^2.
\end{equation}

where we take $d_{ij}$ to be equal to the distances at the beginning of the simulation, such that the sphere structure has vanishing stretching energy (as in \cite{vliegenthart_compression_2011}). In addition to this potential every triangle (found through Delauney triangulation) on the surface of the sphere also has a bending potential\cite{saric_self-assembly_2013}. 

\begin{equation}
E_b = b \sum_{ij} (1-\mathbf{n_i}.\mathbf{n_j})
\end{equation}
where $\mathbf{n_i}$ and $\mathbf{n_j}$ are the normal vectors to neighboring triangles. These are characterized through an energy scale $b$. In order to look at the efficacy of our polymerization schemes, we subject these shells, with different $k_s$ and $b$ to chosen axial extension forces. This gives baselines as to how much force is being generated by a particular polymerization scheme. The results are presented in the supplementary Figure S1.

To measure the propensity of the polymers to induce a deformation, we use the aspect ratio of the elastic shell as a measure, which is defined as the ratio of the longest axis to shortest axis of the ellipse that best fits the elastic shell (see fig. \ref{fig:fig1}). 

Every particle in our system, whether on the shell or in the polymers, interacts with other particles via a modified Weeks-Chandler-Anderson potential (WCA)~\cite{weeks_role_1971}:
\begin{equation} \label{eq:WCA}
\Phi_{\text{WCA}}(r) = 
\begin{cases} 
4\epsilon_{HS} \left[ \left(\frac{\sigma}{\mathbf r}\right)^{12} - \left(\frac{\sigma}{r}\right)^{6} \right] + \epsilon_{HS}, & \text{for } r \leq 2^{1/6}\sigma, \\
0, & \text{for } r > 2^{1/6}\sigma. 
\end{cases}
\end{equation}
which models excluded volume effects. We set $\sigma=1$ and $\epsilon_{HS}=3$ for all particles in all simulations presented here.

\subsection{Toy Model of Polymerization}

The first model we consider is a simple scenario in which $N$ polymers of size $M$ are randomly placed within the interior of a shell. At a rate $R$, a randomly chosen polymer grows by the size of single monomer being added to the end. However, this polymerization only occurs if the added monomer does not overlap with any other monomer.  This model ignores the diffusion and steric interactions of the end joining monomer until it is added. Additionally, features such as nucleation of new polymers are also not included in this model. Roughly, we expect that this model represents the ``best possible world" for polymerization generated forces, in that a free monomer is always avaliable for end joining at a specified rate (in our simulations we choose it such that a monomer is added on average every 2000 simulation timesteps) so long as the there is no obstacle in the way for it to join onto the end.

Additionally, the polymers have a bending potential interaction given by:
\begin{equation}
E_{bp}=\frac{1}{2}\kappa \cos\theta_{ijk}^2
\end{equation}
where $i,j,k$ are a sequence of neighboring monomers. The scale $\kappa$ sets the strength of the bending potential. The persistence length of the polymer scales as $\l_p\sim\kappa$ (see e.g. \cite{zhang_persistence_2019})

\subsection{Modeling the Interactions of Self Assembling Monomers}

In order to model the polymerization of monomers  more accurately, we introduce a coarse grained patchy particle model. Particles are modelled as spheres that can interact via patches on their surface, represented as black or white boxes in Figure~\ref{fig:fig1}A. The number,  location, and binding pattern of patches depends on the particular monomer structure we examine; for example Figure~\ref{fig:fig1}A shows monomers with two patches or four; patches can interact with other specific patches (represented in identical color), and otherwise do not interact. The binding energy between two interacting patches on different monomers is modeled via a modified Kern-Frenkel potential~\cite{rovigatti_how_2018}, where patch $\alpha$ on particle $i$ interacts with patch $\beta$ on particle $j$ using the following potential:
\begin{equation} \label{eq:fullpatch}
\phi_{\text{patch}}(\mathbf r_{ij}, \mathbf{r}_{\alpha,i}, \mathbf{r}_{\beta,j})=V_{SW}( r_{ij})f(\mathbf r_{ij},\mathbf{r}_{\alpha,i},\mathbf{r}_{\beta,j}),
\end{equation}
where 
\begin{equation}\label{eq:squarewell}
V_{SW}(r)=-\epsilon \exp\left[-\frac{1}{2}\left(\frac{r}{\delta}\right)^{10}\right],
\end{equation}
\begin{equation}\label{eq:patch}
f(\mathbf r_{ij},\mathbf{r}_{\alpha,i},\mathbf{r}_{\beta,j}) = 
\begin{cases} 
\cos\left(\frac{\pi \theta_i}{2\theta_m}\right)\cos\left(\frac{\pi \theta_j}{2 \theta_m}\right), &\, \text{if}\, \cos(\theta_i)\ge\cos(\theta_m) \,\text{and}\,\cos(\theta_j)\ge\cos(\theta_m),  \\
0.
\end{cases}
\end{equation}
where $SW$ denotes that the patches have a roughly square-well potential as a function of distance between the centers of mass of the particles $r$. We denote the energy $\epsilon$ as the ``binding energy" of the monomers.

Additionally $\mathbf r_{ij}$ is the vector between particles $i$ and $j$, $\mathbf{r}_{\alpha,i}$ is the vector joining the centre of particle $i$ to patch $\alpha$, $\mathbf{r}_{\beta,j}$ is the vector joining the centre of particle $j$ to patch $\beta$, and the angles $\theta_i$ and $\theta_j$ are the angles to the patch relative to the vector joining the particle centres $\cos \theta_i= \hat{\mathbf{r}}_{ij}.\hat{\mathbf{r}}_{\alpha,i}$. There are three additional key  parameters in the model: $\delta,\theta_m$ and $\epsilon$. Parameter  $\delta$ is a length scale that captures the gives the effective binding distance between patches; $\theta_m$ describes the maximum potential binding angle; and $\epsilon$ is an effective binding energy. In our simulations, we set a small bond length $\delta=1.2\sigma$ in order for patches to only interact with at most one other patch at the same time (where $\sigma$ is the size of a monomer).  We will examine the behavior of the model in a range of $\theta_m$ and $\epsilon$.

\subsection{Simulating the Dynamics of Self Assembling Monomers in Soft Confinement}

We perform molecular dynamics simulations with these potentials subject to a non-equilibrium form of dynamics of adding monomers into the system. Standard models of particle addition and removal from a simulation (for example using the Widom insertion method) would correspond to a grand canonical equilibrium ensemble. We instead imagine a form of particle addition under direct control of the experimenter, who decides the rate $R$ at which particles are released inside the elastic shell  (in practice, this could be done through monomers that are activated or produced within the compartment). This is modelled through particles being added to interior of the elastic shell every $n$ time steps with a probability $n \mathrm{d} t  R$ where $\mathrm{d} t$ is the timestep of the simulation.

Our system is  out of equilibrium because we begin our simulations with a completely empty elastic shell and observe the dynamics of elastic shell deformation as monomers are added at a rate we control. Furthermore, our system is out of equilibrium because as  monomers with large binding energies are added to the system, the system may become trapped in a multitude of metastable states. (We will find that large monomer binding energies are a prerequisite for building polymers sufficiently rigid to deform the compartment.) We will seek means to avoid kinetic traps, by  controlling how monomers interact in space, and how they are released over time. 

We use a Langevin thermostat with a BBK integrator to evolve our system in time~\cite{finkelstein_comparison_2020}. We set a temperature $k_B T=1$. Our measurements of the system will be relative to two fundamental timescales. We can either measure the time in terms of the rate of release $R$, defining $\tau_R=1/R$, such that the timescale $\tau_R$ is the average time required to release one subunit into the system. Another timescale we can use is the diffusive timescale, which is useful as a point of comparison when no monomers are being added to the system or comparing between different rates of addition. We define the diffusive timescale $\tau_D$ as the time it takes to approximately diffuse one subunit diameter ($\sigma$), with $\tau_D=\sigma^2/D$, where $D$ is the diffusion constant in the system, given by the Einstein relation $D=k_b T/3\pi\mu \sigma$ (we set the viscosity $\mu=1$). Our simulations are measured in units of $\tau$ which is defined as $\tau=5235\tau_D$. This timescale corresponds to approximate motion of a diffusive motion over $4r$ where $r$ is the radius of the relaxed shell. This choice of timescale is sufficient for monomer diffusion across the entirety of shell.

For the simulations involving the patchy particles, they also undergo rotational diffusion with a diffusion constant $D_R=k_b T/\pi\mu \sigma^3$, using the integrator discussed in Li et al.~\cite{li_general_2018}.

%%%%%%%%%%%%%%%%%%%%%%%%%%%%%%%%%%%%%%%%%%%%%%%%%%%%%%%%%%%%%%%%%%%%%%%%%%%%%%%
%%%%%%%%%%%%%%%%%%%%%%%%%%%%%%%%%%%%%%%%%%%%%%%%%%%%%%%%%%%%%%%%%%%%%%%%%%%%%%%
\section{Results}
%%%%%%%%%%%%%%%%%%%%%%%%%%%%%%%%%%%%%%%%%%%%%%%%%%%%%%%%%%%%%%%%%%%%%%%%%%%%%%%
%%%%%%%%%%%%%%%%%%%%%%%%%%%%%%%%%%%%%%%%%%%%%%%%%%%%%%%%%%%%%%%%%%%%%%%%%%%%%%%

\subsection{A Toy Model of Polymerization Inside an Elastic Shell Reveals Spontaneous Bundling Leading to Shell Deformation}
%%%%%%%%%%%%%%%%%%%%%%%%%%%%%%%%%%%%%%%%%%%%%%%%%%%%%%%%%%%%%%%%%%%%%%%%%%%%%%%

\begin{figure}[h]
    \begin{center}
    \includegraphics[width=180mm]{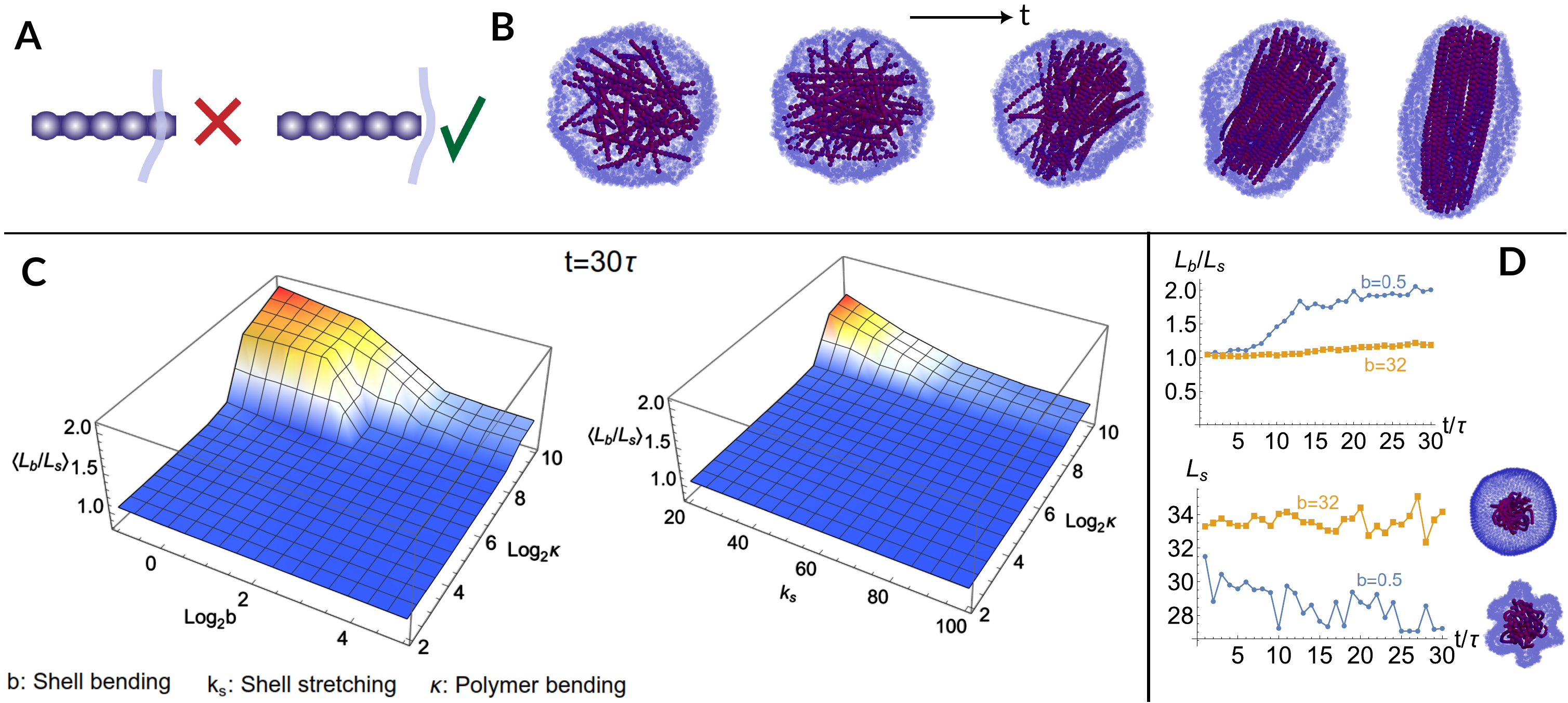}
    \caption{Toy model of polymerization within a soft compartment. A) Monomers are added to a polymer with some rate $R$ so long as the added monomer does not overlap with with other particles. B) For polymers with high persistence lengths, the addition of monomers eventually causes the bundling of the separate polymers. Once bundled, they continue to grow and push the shell into deformed shapes. C) The influence of the elastic parameters in our problem on the average deformation (major to minor axis ratio) after a time $t=30\tau$. Shell stiffness and bending parameters appear to have similar effects, while very stiff polymers are required for deformation. D) Major to minor axes ratio and minor axes size for different shell bending strengths. High shell bending potentials resist buckling, which reduces the deformation ratio. }
    \label{fig:figTM} %
    \end{center}
    \end{figure}

A full dynamical model of polymerization within soft confinement involves the confluence of two factors. The first is the material properties of the polymers and the elastic shell, and the second is the process of monomer addition and removal. In order to understand fully which material elastic properties are necessary to enable axial deformation, we first introduce a toy model for polymerizing subunits that neglects some of the physical details (see Methods). Here nucleation of new polymers is not modeled, while existing polymers grow by single monomer addition at rate $R$, provided that an added monomer does not overlap with any other monomer or with the shell as illustrated in Figure \ref{fig:figTM}A. Diffusion and steric interactions of the added monomer are ignored until it is added. Thanks to its simplicity, this toy model is useful to gain general intuition about the physical processes important for polymerization inside soft compartments. 

    Our first important observation is that growing polymers in a soft compartment have a tendency to bundle, as illustrated in Figure~\ref{fig:figTM}B. This behavior appears linked to soft compartments, as similar models of growing polymers inside a hard compartment show no bundling behavior (supplementary Figure S2). The formation of bundles allows the polymers to ``cooperate" to continue growing together, leading to elastic shells with large axial deformations, by forcing all the polymers to push in the same direction. We systematically evaluated axial deformation for different values of the parameters governing the elasticity of both the semiflexible polymers and of the elastic shell, shown in Figure~\ref{fig:figTM}C. The shell parameters $k_s$ and $b$ seem to have similar effects on the propensity of the polymers to deform the compartment. As one would expect, increasing the strength of either the shell stretching or bending parameters makes it harder for the polymers to stretch the compartment.  Additionally, the persistence length of polymers has to be high in order to observe any deformation-multiples of the total length of the compartment. If the persistence length is insufficient, polymers merely wrap around the interior of the compartment. Examples of the temporal evolution of major and minor axes for different shell parameters are in Figure~\ref{fig:figTM}D.

    While one could theoretically imagine situations in which growing polymers do not tend to align and push in the same direction, we do not observe this in our simulations. For example, one can imagine the polymers pushing on the shell isotropically. In practice, making the elastic forces sufficiently weak for this to occur can also lead to the polymers "puncturing" the shell, thus growing continually without leading to further force generation. We shall have more to say on the bundling transition in the next sections.

%%%%%%%%%%%%%%%%%%%%%%%%%%%%%%%%%%%%%%%%%%%%%%%%%%%%%%%%%%%%%%%%%%%%%%%%%%%%%%%
    \subsection{Self Assembling Polymer Structures Require Slow Monomer Release for Axial Deformation}
%%%%%%%%%%%%%%%%%%%%%%%%%%%%%%%%%%%%%%%%%%%%%%%%%%%%%%%%%%%%%%%%%%%%%%%%%%%%%%%

    We next examine compartment deformation with a more detailed model for the monomer addition process. The toy model ignored the fact that free monomers may generally not be present near the end of a growing filament, that  monomers should have the correct orientation to be added to the growing polymer, and that new polymers may be nucleated when there is a high abundance of monomers. To address these shortcomings, monomers in the detailed model undergo translational and rotational diffusion. When a compatible binding site is found, the monomer becomes part of larger structure. Monomers are supplied into the interior of the shell at some rate $R$. We anticipated that this model would not achieve axial deformations comparable to what  observed in the toy model, due to the reduced monomer availability near the growing end of the polymer as shells fluctuate. However, an adjustable monomer supply rate $R$ may serve as a "molecular program" to achieve different shell deformation outcomes. In contrast to the toy model, we now vary only the local binding parameters of the polymer, as opposed to the bending moduli of the structures directly through a bending potential. These parameters are $\epsilon$ and $\theta_m$, which are the depth and angular width of the Kern-Frenkel potential, respectively (further details can be found in the Methods). The persistence length of these structures in free space can nevertheless be measured (supplementary Figure S3). 

    \begin{figure}[h]
        \begin{center}
        \includegraphics[width=180mm]{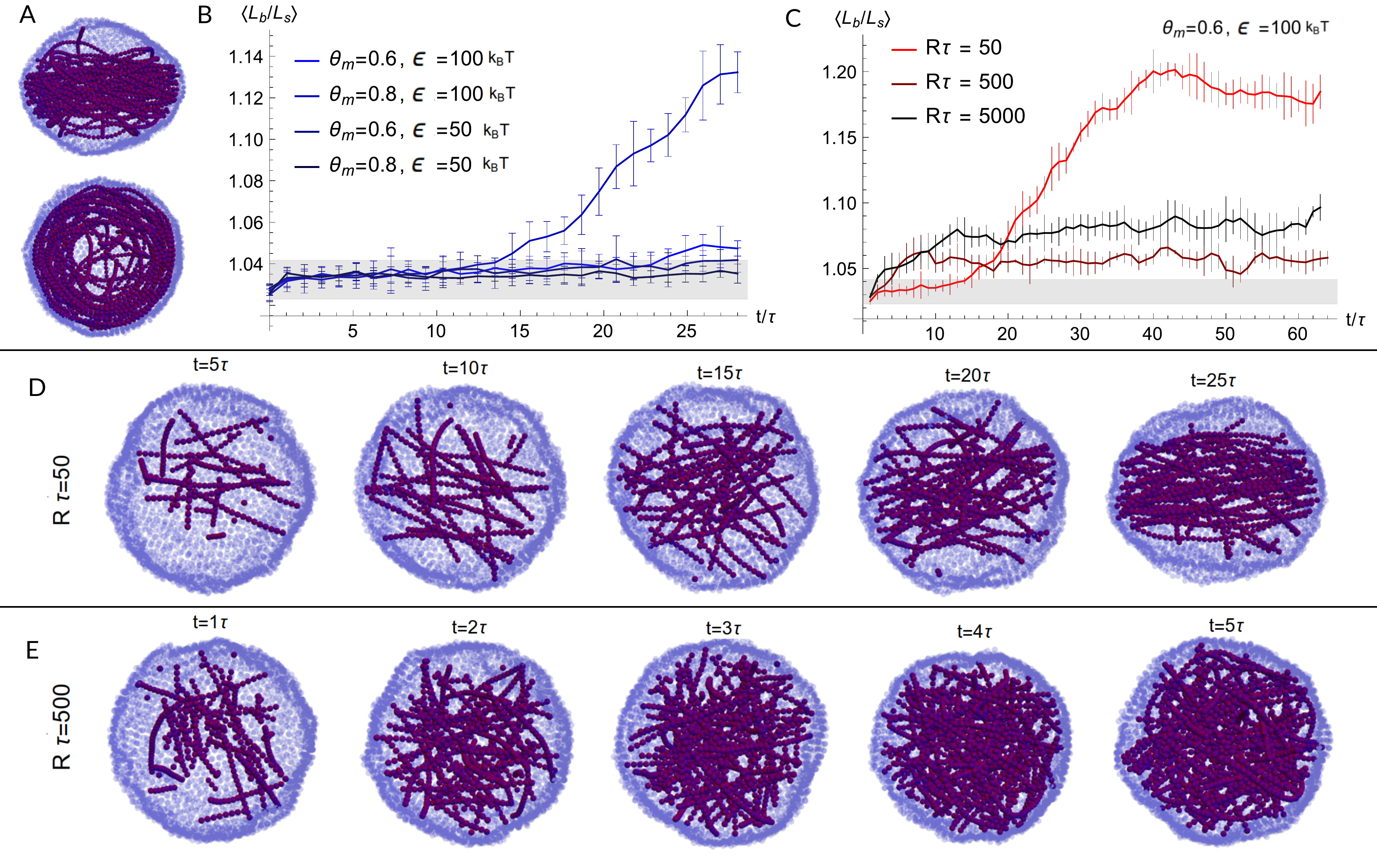}
        \caption{ Self assembly of polymeric structures inside soft shells A) For high persistence lengths, bundles form which deform the shell, whereas for low persistence lengths the polymers form rings around the interior. B) Only high persistence length polymers (reflected in these microscopic parameters) are capable of deforming the shell much beyond its equilibrium fluctuations (gray band) (averaged over 5 different simulations) C) Releasing monomers at different rates can have significant effects on measured deformation. If monomers are released into the interior of the shell too quickly, the total capacity of the polymers to induce deformations goes down significantly. (averaged over 5 different simulations) D) Time evolution of the case where monomers are released slowly, leading to bundles. E) Time evolution of case where monomers are released quickly, leading to bent polymers. }
        \label{fig:figSA} %
        \end{center}
        \end{figure}

        Figure~\ref{fig:figSA} illustrates that our simulations lead to various structures, which can be compared to experiments in which soft polymers are assembled in confinement. For example, the formation of "rings" that snake around the edge of the elastic shell shown in panel A, is  consistent with experimental observations of ring-like structures formed by actin, microtubules, and artificial DNA polymers assembling in elastic shells or droplets~\cite{limozin2002polymorphism,baumann2014motor,agarwal2021dynamic}.  Buckling of the polymers leads to states that make it difficult to induce significant axial forces, as new monomers are not added to the end of a growing polymer brush that is pushing against the edge of the elastic shell, as is seen in panel B. Like in our toy model, when the microscopic parameters are associated with large persistence lengths, the polymers form bundles that lead to axial deformation. As one decreases the energy of binding (which controls the persistence length) the polymerization does not lead to any deformation greater than what is observed for the same elastic shell under thermal fluctuations. However, persistence length is not the only important parameter: variations in the monomer supply rate $R$ can dramatically affect deformation. Panel C shows that deformation is minimal if the monomers are released at a sufficiently fast rate, while larger deformations occur when monomers are supplied slowly. This difference may stem from the fact that, in contrast to the toy model, these simulations do not assume a fixed number of polymers. If the supply rate is too high (panel E), many small polymers are formed, leading to small deformations of the vesicle on fast timescales, which however saturate quickly. In contrast, when monomers are slowly supplied into the interior a larger total possible deformation can be achieved, although this requires a much longer timescale (panel D). This implies a tradeoff between the speed and magnitude of forces that can be generated via polymerization, which corresponds to a tradeoff between power and total work that can be delivered in the system. We note that the difference between $R\tau=500$ and $R\tau=5000$ is small, but $R\tau=5000$ is somewhat higher. We do not consider this difference significant, however.

        Full phase diagrams for the effects of these microscopic parameters on shell deformation are to be found in the supplementary Figures S4 and S5. These phase diagrams are qualitatively consistent with the behavior of the toy model.

%%%%%%%%%%%%%%%%%%%%%%%%%%%%%%%%%%%%%%%%%%%%%%%%%%%%%%%%%%%%%%%%%%%%%%%%%%%%%%%
    \subsection{Mechanisms of Shell Deformation}
%%%%%%%%%%%%%%%%%%%%%%%%%%%%%%%%%%%%%%%%%%%%%%%%%%%%%%%%%%%%%%%%%%%%%%%%%%%%%%%
    Both the toy model and the detailed  model show the potential for growing polymers to deform the elastic shell along one axis, although why this occurs is less clear. One could plausibly imagine alternative scenarios where the  filaments only grow isotropically,  rather than choosing a particular direction that induces axial deformation. Are there specific mechanisms that promote directional force generation? 
    
    To address this question,  we quantified several characteristics of the simulations presented in Figures~\ref{fig:figTM} and~\ref{fig:figSA}.  We measured the size of the largest polymer in terms of the number of incorporated monomers $N$, the average size of the ten largest polymers in the system, $<N>$, and the \EF{``normalized extension''} of polymers, measured as their end to end distance over the number of monomers, \EF{$R_{G}/N$}. Finally, we assessed the tendency of growing polymers to bundle. Qualitatively, the images suggest that polymers bundle together and then grow directionally against the shell. To clarify this process, we introduce and measure a "bundling score":
    \begin{equation}
    B_{ij} = \mathbf{v}_i.\mathbf{v}_j
    \end{equation}
    where $\mathbf{v}_i$ is the average direction of all the bond vectors making up the polymer $i$. The average direction is computed as $\mathbf v= \frac{1}{N-1}\sum_{j=1}^{N-1}(\mathbf r_{l+1}-\mathbf r_{l})$ where $\mathbf r_j$ is the position vector of monomer $l$ and $N$ is the total number of monomers on the polymer.
     For each pair $(i,j)$ of polymers in the system, we can calculate the average bundling score, weighted by the distance between the polymers. This weighted bundling score will be equal to $0.5$ if the polymers aren't aligned at all, and to $1$ for perfect alignment.

    \begin{figure}[h]
        \begin{center}
        \includegraphics[width=180mm]{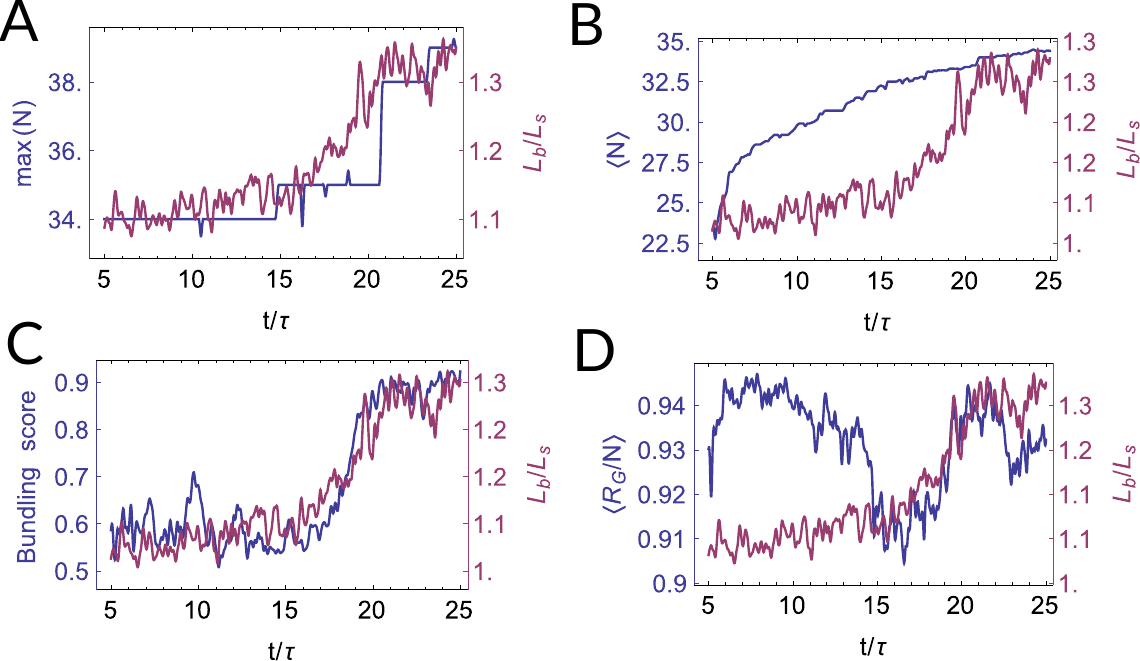}
        \caption{ Selected microscopic details of the polymers inside the shell during the time evolution of the system. All plots overlaid with the total deformation (purple). A) The maximum polymer size (in monomers) measured in the system over size. This quantity remains relatively stable. B) The average size of the 10 largest monomers. This quantity grows but slowly. C) Bundling score (see text) over time. The deformation is highly correlated to the bundling. D) The average over the largest 10 polymers of the end to end distance over number of monomers. The polymers gradually get more bent as the simulation proceeds, before extending again as the shell extends.}
        \label{fig:figMI} %
        \end{center}
        \end{figure}

     All these features are plotted in  Figure \ref{fig:figMI}, side by side with the shell deformation $L_b/L_s$. An increase in the deformation of the shell is not strictly tied to the size of the largest polymer or to the average size of the largest polymers (panels A and B). However, when looking at the total bundling score (panel C) and the extension of the polymers (panel D), we find that an increase of these quantities appears to be more closely correlated with an increase in shell deformation.  This suggests the following candidate mechanism to explain deformation of an elastic shell by semi-flexible polymers: as the polymers grow, they tend to get more ``bent", as suggested by a decrease of their normalized extension in panel D). This creates a store of elastic energy in the polymers, but they are individually not strong enough to deform the shell. Through a random fluctuation of the shell, one polymer extends against the shell, producing a preferred axis that is slightly longer than the other axes in the system. As the other polymers undergo rotational motion, they eventually find this longer axis and are able to extend more in this direction, becoming  aligned with the original polymer. The combination of many polymers pushing in this direction is enough to lead to significant deformation of the shell along this axis. One then sees an increase in the polymer normalized extension and in the total bundling score. We note that this mechanism is only plausible when the monomers are released slowly. If monomers are released fast, many short polymers nucleate quickly and then end-join forming polymers which are too long to find any preferential direction to stretch. 

    Finally, the bundling transition we observe in these systems is contingent on the shell being soft. Simulations of our toy model in which polymers grow within a hard shell (i.e., there are no shell fluctuations) do not show any bundling transition, as shown in supplementary Figure S2. We can conclude that bundling is favored by cooperative effects that arise from the interactions of the shell with the growing polymers. The feedback between growing structures and the responsive barrier they push against may be key to inducing a rich set of behaviors and states.  
%%%%%%%%%%%%%%%%%%%%%%%%%%%%%%%%%%%%%%%%%%%%%%%%%%%%%%%%%%%%%%%%%%%%%%%%%%%%%%%
\subsection{Crosslinking Particles Have Mixed Effects on Deformation}
%%%%%%%%%%%%%%%%%%%%%%%%%%%%%%%%%%%%%%%%%%%%%%%%%%%%%%%%%%%%%%%%%%%%%%%%%%%%%%%

We next ask whether shell deformation can be enhanced by the addition of multivalent monomers serving as polymer crosslinkers. Because bundling emerged as a key strategy to produce directional forces, it stands to reason that by promoting polymer interactions through organizing particles one could improve force generation, even with polymers which by themselves are not sufficiently stiff. We thus introduce crosslinkers in our simulation, and we do so dynamically like for regular monomers: any particle in the pool is supplied into the elastic shell at a rate proportional to their total proportion (i.e., if 30\% of monomers are multivalent, when a new particle is released it will be a multivalent particle on average 30\% of the time). This approach hides a deceptive form of complexity, that is the properties of the system will depend on both the \textit{structure} of crosslinkers as well as on \textit{how} they are introduced.  

\clearpage

   \begin{figure}[h]
        \begin{center}
        \includegraphics[width=180mm]{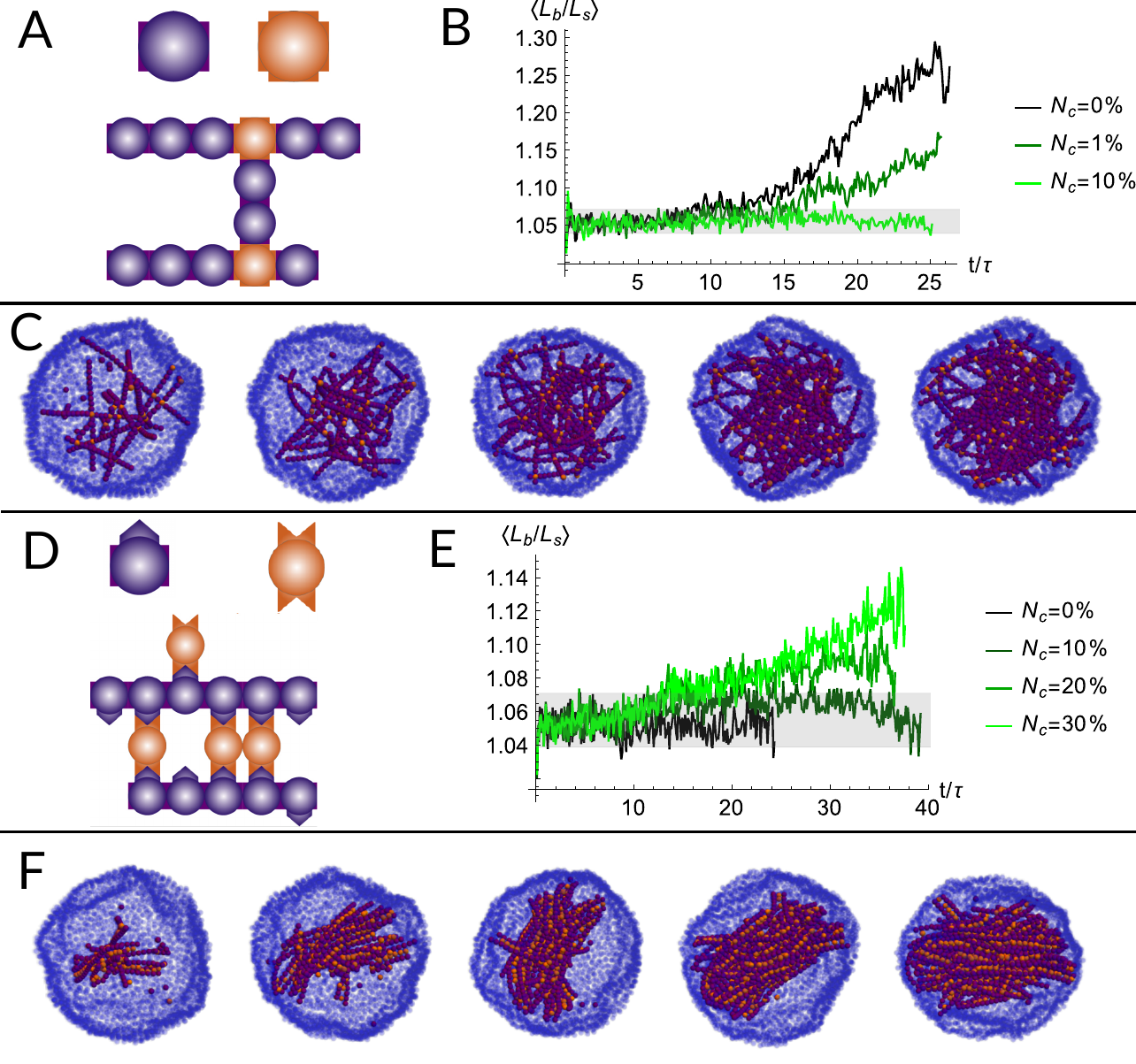}
        \caption{How the introduction of multivalent particles changes the performance of a molecular system to generate deformation. A) We include a fraction of monomers operating as multivalent particles that can act as crosslinkers. B) Crosslinking tends to impede the generation of axial forces, even with only 1\% of crosslinker monomers; here  $\epsilon=100,\theta_m=0.6,R\tau=50$. C) The detrimental effect of crosslinking is caused by isotropic growth of the network. D) A different scheme where every monomer has a potential binding site on its surface for the binding of over crosslinking monomers. We indicate the restrictions on potential binding by coloring the sites red and blue, where red and blue can interact with each other but not themselves. E) This scheme can lead to axial forces even for monomers which wouldn't ordinarily deform the compartment $\epsilon=50,\theta_m=0.8,R\tau=50$. F) The crosslinkers force the growing polymer network to adopt a bundled structure from the beginning of the simulation. All simulations performed for shell parameters $k_s=20,b=1$.}
        \label{fig:figXL} %
        \end{center}
        \end{figure}
\clearpage

We begin by including a fraction of crosslinker monomers (valency 4) in a pool of monomers (valency 2) that were found to spontaneously bundle and deform an elastic shell (Figure~\ref{fig:figMI}). The simulations in Figure~\ref{fig:figXL} demonstrate that these crosslinkers actually hinder the generation of forces (panel B), as they lead to a more isotropic network that isn't able to push effectively in any direction. Even 1\% of multivalent particles is sufficient to reduce deformation by about 10\%.  When the crosslinker fraction is increased to 10\%, the ability of the network to lead to axial deformations is destroyed. 

We then asked if the opposite case is possible, that is whether we could find particular types of crosslinkers leading to shell deformation in a monomer system that ordinarily fails to do so. We examined several different crosslinking schemes, and the most effective we found is shown in Figure~\ref{fig:figXL}D. In this case, every monomer has one potential binding site for crosslinkers, and crosslinkers can only introduce a link to different polymers while they cannot bind to each other. This scheme can lead to polymer networks that deform the shell, while non-crosslinked polymers are not stiff enough to produce deformation. In the presence of crosslinkers polymers become quickly bundled as they are growing; the total proportion of monomers to crosslinkers is still a very important parameter.  In  supplementary Figures S6 and S7 we show other variants of this scheme, that consider the presence on each monomer of 2 potential binding sites for crosslinkers, and the possibility of crosslinkers to bind to one another. These schemes do not tend to change our qualitative conclusions.

%%%%%%%%%%%%%%%%%%%%%%%%%%%%%%%%%%%%%%%%%%%%%%%%%%%%%%%%%%%%%%%%%%%%%%%%%%%%%%%
%%%%%%%%%%%%%%%%%%%%%%%%%%%%%%%%%%%%%%%%%%%%%%%%%%%%%%%%%%%%%%%%%%%%%%%%%%%%%%%
\section{Discussion and Conclusion}
%%%%%%%%%%%%%%%%%%%%%%%%%%%%%%%%%%%%%%%%%%%%%%%%%%%%%%%%%%%%%%%%%%%%%%%%%%%%%%%
%%%%%%%%%%%%%%%%%%%%%%%%%%%%%%%%%%%%%%%%%%%%%%%%%%%%%%%%%%%%%%%%%%%%%%%%%%%%%%%

In this study, we were able to find conditions under which the passive assembly of polymers inside a soft elastic shell produces forces sufficient to induce deformation of the shell.  The generation of the forces was found to depend sensitively on the underlying mechanical properties of the polymer and the rate at which monomers are released into the elastic shell. A slow, constant release rate of monomers promotes bundling and therefore axial deformation, which is favored when the shell fluctuations permit local monomer addition. The details of how monomers interact can lead to differences in force generation, and the inclusion of  crosslinker particles can promote either directional forces or isotropic forces, depending on the crosslinking rules.

Classical works of force generation by polymers have identified different mechanisms by which ``polymerization"  can generate forces~\cite{theriot_polymerization_2000}. In particular, ideas from statistical physics led to the definitions of  "Brownian ratchet" mechanism of force generation~\cite{peskin_cellular_1993}, as well as of  ``elastic Brownian ratchet"~\cite{mogilner_cell_1996}. Briefly, the Brownian ratchet can generate force by adding a monomer to the end of a growing filament as the load fluctuates thermally. In the elastic Brownian ratchet, the filaments ``bend", and the elastic restoring force of the bent filament can push the load. Our simulations suggest that the force generation mechanism in our system doesn't neatly fit into either category, but is more similar to the elastic Brownian ratchet: in the case where the polymers are sufficiently stiff to lead to deformation of the membrane, they appear to do this by bundling together and straightening themselves.

Our study assumes that the rate of release of monomers into the elastic shell is a quantity under experimental control. Thermodynamic models of polymerization forces state that the force generation capacity of polymerization is given by\cite{dmitrieff_amplification_2016}:
\begin{equation}
    F_{max} = k_b T \ln(C/C^{*})/\sigma
\end{equation}
where $C$ is the concentration of free monomers, $C*$ is a critical concentration required for polymerization to occur, and $\sigma$ is the length of a single monomer. By fixing the rate of release, we are in effect keeping the pool of available monomers fixed at a particular value $C$, which does not diminish as new monomers are added to the system. Nor are monomers dynamically removed from inside the shell once they are added into it. Our model could be readily extended to account for these facts, or even to simulate equilibrium system with a specified chemical potential. Here, however, our focus was on which molecular details and timed release of subunits could lead to the generation of forces. Experimental realizations of using polymers to produce forces dynamically would therefore require special control over quantities such as the concentration of available monomers, such as activating them via light.

Bundling being a key aspect of force generation in these elastic shells, we found that multivalent particles can be helpful in forcing the polymers to bundle along a particular direction and thus lead to improved deformation, even for soft polymers that by themselves would not be rigid enough to deform the elastic shell. However, the details of the crosslinking matter. Introducing few monomers that work as crosslinkers can make the subsequent polymer network  isotropic. On the opposite side of the spectrum, complex assemblies with many types of monomers can lead to a broad landscape of possible misfolded targets making it harder to achieve the desired target. We found that crosslinkers working  as a monomer variant that can't polymerize but can bind anywhere on a polymer are the most successful in generating bundled structures, leading to the generation of axial force. Such a design ensures that the crosslinks can act as a ``zip" between neighboring polymers. This suggests a high degree of crosslinking is necessary to overcome the lack of sufficient rigidity of individual polymers to perform work?. However, performance degrades under an excess of crosslinkers, indicating that there is an optimal number of crosslinkers required for maximizing force generation. %Experimental validations of similar crosslinking schemes~\cite{zhang2023building} produce DNA filament bundles, and their gradual assembly in confinement may lead to compartment deformation.
One of the main challenges with the problem as defined is that the possible space of assembly pathways and types of monomers is essentially limitless. While we have focused on a few (what we consider to be) sensible molecular programs, the potential for other programs to perform much better still exists but may have been missed. Our explorations in varying release rates over time were rudimentary, but seemed to indicate that keeping release rates held fixed was the best for force generation. This does not, however, constitute a proof of this. Our mechanism of introduction of monomers into the elastic shell assumes an initially empty sphere into which monomers are added, rather than the perhaps more realistic case where hard-sphere monomers already exist inside the elastic shell and are activated in time. Our choice expedites simulation time and we do not anticipate that great differences would arise were the alternative to be considered, but the restriction of available space could conceivably help or harm any particular force generation program (we simulate this possibility in supplementary Figure 8 and do not observe differences from our main results). Additionally, our work uses interaction rates which are essentially irreversible on the timescale of the simulation length. In other words, once two monomers have bound together, they do not unbind. This allows us to ignore an additional parameter in our simulations, however experimental polymer systems exhibit non-negligible dissociation rates, which have an influence on the kinetics and equilibrium states.

\section{Acknowledgements}

This work was supported by the U.S. Department of Energy, Office of Science, Basic Energy Sciences under Award \# DE-SC0010595. 

\bibliography{references}

\providecommand{\latin}[1]{#1}
\makeatletter
\providecommand{\doi}
  {\begingroup\let\do\@makeother\dospecials
  \catcode`\{=1 \catcode`\}=2 \doi@aux}
\providecommand{\doi@aux}[1]{\endgroup\texttt{#1}}
\makeatother
\providecommand*\mcitethebibliography{\thebibliography}
\csname @ifundefined\endcsname{endmcitethebibliography}
  {\let\endmcitethebibliography\endthebibliography}{}
\begin{mcitethebibliography}{48}
\providecommand*\natexlab[1]{#1}
\providecommand*\mciteSetBstSublistMode[1]{}
\providecommand*\mciteSetBstMaxWidthForm[2]{}
\providecommand*\mciteBstWouldAddEndPuncttrue
  {\def\EndOfBibitem{\unskip.}}
\providecommand*\mciteBstWouldAddEndPunctfalse
  {\let\EndOfBibitem\relax}
\providecommand*\mciteSetBstMidEndSepPunct[3]{}
\providecommand*\mciteSetBstSublistLabelBeginEnd[3]{}
\providecommand*\EndOfBibitem{}
\mciteSetBstSublistMode{f}
\mciteSetBstMaxWidthForm{subitem}{(\alph{mcitesubitemcount})}
\mciteSetBstSublistLabelBeginEnd
  {\mcitemaxwidthsubitemform\space}
  {\relax}
  {\relax}

\bibitem[Krishna \latin{et~al.}(2024)Krishna, Wang, Ahern, Sturmfels,
  Venkatesh, Kalvet, Lee, Morey-Burrows, Anishchenko, Humphreys, \latin{et~al.}
  others]{krishna2024generalized}
Krishna,~R.; Wang,~J.; Ahern,~W.; Sturmfels,~P.; Venkatesh,~P.; Kalvet,~I.;
  Lee,~G.~R.; Morey-Burrows,~F.~S.; Anishchenko,~I.; Humphreys,~I.~R.,
  \latin{et~al.}  Generalized biomolecular modeling and design with RoseTTAFold
  All-Atom. \emph{Science} \textbf{2024}, eadl2528\relax
\mciteBstWouldAddEndPuncttrue
\mciteSetBstMidEndSepPunct{\mcitedefaultmidpunct}
{\mcitedefaultendpunct}{\mcitedefaultseppunct}\relax
\EndOfBibitem
\bibitem[Liu \latin{et~al.}(2022)Liu, Appel, Ashby, Baker, Franco, Gu, Haynes,
  Joshi, Kloxin, Kouwer, \latin{et~al.} others]{liu2022living}
Liu,~A.~P.; Appel,~E.~A.; Ashby,~P.~D.; Baker,~B.~M.; Franco,~E.; Gu,~L.;
  Haynes,~K.; Joshi,~N.~S.; Kloxin,~A.~M.; Kouwer,~P.~H., \latin{et~al.}  The
  living interface between synthetic biology and biomaterial design.
  \emph{Nature materials} \textbf{2022}, \emph{21}, 390--397\relax
\mciteBstWouldAddEndPuncttrue
\mciteSetBstMidEndSepPunct{\mcitedefaultmidpunct}
{\mcitedefaultendpunct}{\mcitedefaultseppunct}\relax
\EndOfBibitem
\bibitem[Green \latin{et~al.}(2019)Green, Subramanian, Mardanlou, Kim, Hariadi,
  and Franco]{green2019autonomous}
Green,~L.~N.; Subramanian,~H.~K.; Mardanlou,~V.; Kim,~J.; Hariadi,~R.~F.;
  Franco,~E. Autonomous dynamic control of {DNA} nanostructure self-assembly.
  \emph{Nature chemistry} \textbf{2019}, \emph{11}, 510--520\relax
\mciteBstWouldAddEndPuncttrue
\mciteSetBstMidEndSepPunct{\mcitedefaultmidpunct}
{\mcitedefaultendpunct}{\mcitedefaultseppunct}\relax
\EndOfBibitem
\bibitem[Agarwal and Franco(2019)Agarwal, and Franco]{agarwal2019enzyme}
Agarwal,~S.; Franco,~E. Enzyme-driven assembly and disassembly of hybrid
  {DNA}--{RNA} nanotubes. \emph{Journal of the American Chemical Society}
  \textbf{2019}, \emph{141}, 7831--7841\relax
\mciteBstWouldAddEndPuncttrue
\mciteSetBstMidEndSepPunct{\mcitedefaultmidpunct}
{\mcitedefaultendpunct}{\mcitedefaultseppunct}\relax
\EndOfBibitem
\bibitem[Jorgenson \latin{et~al.}(2017)Jorgenson, Mohammed, Agrawal, and
  Schulman]{jorgenson2017self}
Jorgenson,~T.~D.; Mohammed,~A.~M.; Agrawal,~D.~K.; Schulman,~R. Self-assembly
  of hierarchical {DNA} nanotube architectures with well-defined geometries.
  \emph{ACS nano} \textbf{2017}, \emph{11}, 1927--1936\relax
\mciteBstWouldAddEndPuncttrue
\mciteSetBstMidEndSepPunct{\mcitedefaultmidpunct}
{\mcitedefaultendpunct}{\mcitedefaultseppunct}\relax
\EndOfBibitem
\bibitem[Agrawal \latin{et~al.}(2017)Agrawal, Jiang, Reinhart, Mohammed,
  Jorgenson, and Schulman]{agrawal2017terminating}
Agrawal,~D.~K.; Jiang,~R.; Reinhart,~S.; Mohammed,~A.~M.; Jorgenson,~T.~D.;
  Schulman,~R. Terminating {DNA} tile assembly with nanostructured caps.
  \emph{ACS nano} \textbf{2017}, \emph{11}, 9770--9779\relax
\mciteBstWouldAddEndPuncttrue
\mciteSetBstMidEndSepPunct{\mcitedefaultmidpunct}
{\mcitedefaultendpunct}{\mcitedefaultseppunct}\relax
\EndOfBibitem
\bibitem[Zhang \latin{et~al.}(2023)Zhang, Yang, Wang, and
  Ke]{zhang2023building}
Zhang,~Y.; Yang,~D.; Wang,~P.; Ke,~Y. Building Large {DNA} Bundles via
  Controlled Hierarchical Assembly of DNA Tubes. \emph{ACS nano} \textbf{2023},
  \emph{17}, 10486--10495\relax
\mciteBstWouldAddEndPuncttrue
\mciteSetBstMidEndSepPunct{\mcitedefaultmidpunct}
{\mcitedefaultendpunct}{\mcitedefaultseppunct}\relax
\EndOfBibitem
\bibitem[Mitchell \latin{et~al.}(2004)Mitchell, Harris, Malo, Bath, and
  Turberfield]{mitchell2004self}
Mitchell,~J.~C.; Harris,~J.~R.; Malo,~J.; Bath,~J.; Turberfield,~A.~J.
  Self-assembly of chiral {DNA} nanotubes. \emph{Journal of the American
  Chemical Society} \textbf{2004}, \emph{126}, 16342--16343\relax
\mciteBstWouldAddEndPuncttrue
\mciteSetBstMidEndSepPunct{\mcitedefaultmidpunct}
{\mcitedefaultendpunct}{\mcitedefaultseppunct}\relax
\EndOfBibitem
\bibitem[Fletcher and Theriot(2004)Fletcher, and
  Theriot]{fletcher_introduction_2004}
Fletcher,~D.~A.; Theriot,~J.~A. An introduction to cell motility for the
  physical scientist. \emph{Physical Biology} \textbf{2004}, \emph{1}, T1\relax
\mciteBstWouldAddEndPuncttrue
\mciteSetBstMidEndSepPunct{\mcitedefaultmidpunct}
{\mcitedefaultendpunct}{\mcitedefaultseppunct}\relax
\EndOfBibitem
\bibitem[Simon \latin{et~al.}(2019)Simon, Kusters, Caorsi, Allard, Abou-Ghali,
  Manzi, Di~Cicco, Lévy, Lenz, Joanny, Campillo, Plastino, Sens, and
  Sykes]{simon_actin_2019}
Simon,~C.; Kusters,~R.; Caorsi,~V.; Allard,~A.; Abou-Ghali,~M.; Manzi,~J.;
  Di~Cicco,~A.; Lévy,~D.; Lenz,~M.; Joanny,~J.-F.; Campillo,~C.; Plastino,~J.;
  Sens,~P.; Sykes,~C. Actin dynamics drive cell-like membrane deformation.
  \emph{Nature Physics} \textbf{2019}, \emph{15}, 602--609\relax
\mciteBstWouldAddEndPuncttrue
\mciteSetBstMidEndSepPunct{\mcitedefaultmidpunct}
{\mcitedefaultendpunct}{\mcitedefaultseppunct}\relax
\EndOfBibitem
\bibitem[Mogilner and Oster(2003)Mogilner, and Oster]{mogilner_force_2003}
Mogilner,~A.; Oster,~G. Force {Generation} by {Actin} {Polymerization} {II}:
  {The} {Elastic} {Ratchet} and {Tethered} {Filaments}. \emph{Biophysical
  Journal} \textbf{2003}, \emph{84}, 1591--1605\relax
\mciteBstWouldAddEndPuncttrue
\mciteSetBstMidEndSepPunct{\mcitedefaultmidpunct}
{\mcitedefaultendpunct}{\mcitedefaultseppunct}\relax
\EndOfBibitem
\bibitem[Dominguez and Holmes(2011)Dominguez, and Holmes]{dominguez_actin_2011}
Dominguez,~R.; Holmes,~K.~C. Actin {Structure} and {Function}. \emph{Annual
  Review of Biophysics} \textbf{2011}, \emph{40}, 169--186, \_eprint:
  https://doi.org/10.1146/annurev-biophys-042910-155359\relax
\mciteBstWouldAddEndPuncttrue
\mciteSetBstMidEndSepPunct{\mcitedefaultmidpunct}
{\mcitedefaultendpunct}{\mcitedefaultseppunct}\relax
\EndOfBibitem
\bibitem[Rottner \latin{et~al.}(2017)Rottner, Faix, Bogdan, Linder, and
  Kerkhoff]{rottner_actin_2017}
Rottner,~K.; Faix,~J.; Bogdan,~S.; Linder,~S.; Kerkhoff,~E. Actin assembly
  mechanisms at a glance. \emph{Journal of Cell Science} \textbf{2017},
  \emph{130}, 3427--3435\relax
\mciteBstWouldAddEndPuncttrue
\mciteSetBstMidEndSepPunct{\mcitedefaultmidpunct}
{\mcitedefaultendpunct}{\mcitedefaultseppunct}\relax
\EndOfBibitem
\bibitem[Lieleg \latin{et~al.}(2010)Lieleg, Claessens, and
  Bausch]{lieleg_structure_2010}
Lieleg,~O.; Claessens,~M. M. A.~E.; Bausch,~A.~R. Structure and dynamics of
  cross-linked actin networks. \emph{Soft Matter} \textbf{2010}, \emph{6},
  218--225, Publisher: The Royal Society of Chemistry\relax
\mciteBstWouldAddEndPuncttrue
\mciteSetBstMidEndSepPunct{\mcitedefaultmidpunct}
{\mcitedefaultendpunct}{\mcitedefaultseppunct}\relax
\EndOfBibitem
\bibitem[Limozin \latin{et~al.}(2003)Limozin, Bärmann, and
  Sackmann]{limozin_organization_2003}
Limozin,~L.; Bärmann,~M.; Sackmann,~E. On the organization of self-assembled
  actin networks in giant vesicles. \emph{The European Physical Journal E}
  \textbf{2003}, \emph{10}, 319--330\relax
\mciteBstWouldAddEndPuncttrue
\mciteSetBstMidEndSepPunct{\mcitedefaultmidpunct}
{\mcitedefaultendpunct}{\mcitedefaultseppunct}\relax
\EndOfBibitem
\bibitem[Tsai \latin{et~al.}(2014)Tsai, Roth, Dogterom, and
  Koenderink]{tsai_chapter_2014}
Tsai,~F.-C.; Roth,~S.; Dogterom,~M.; Koenderink,~G.~H. In \emph{Advances in
  {Planar} {Lipid} {Bilayers} and {Liposomes}}; Iglič,~A., Kulkarni,~C.~V.,
  Eds.; Academic Press, 2014; Vol.~19; pp 139--173\relax
\mciteBstWouldAddEndPuncttrue
\mciteSetBstMidEndSepPunct{\mcitedefaultmidpunct}
{\mcitedefaultendpunct}{\mcitedefaultseppunct}\relax
\EndOfBibitem
\bibitem[Baldauf \latin{et~al.}(2022)Baldauf, Van~Buren, Fanalista, and
  Koenderink]{baldauf2022actomyosin}
Baldauf,~L.; Van~Buren,~L.; Fanalista,~F.; Koenderink,~G.~H. Actomyosin-driven
  division of a synthetic cell. \emph{ACS Synthetic Biology} \textbf{2022},
  \emph{11}, 3120--3133\relax
\mciteBstWouldAddEndPuncttrue
\mciteSetBstMidEndSepPunct{\mcitedefaultmidpunct}
{\mcitedefaultendpunct}{\mcitedefaultseppunct}\relax
\EndOfBibitem
\bibitem[Litschel \latin{et~al.}(2021)Litschel, Kelley, Holz, Adeli~Koudehi,
  Vogel, Burbaum, Mizuno, Vavylonis, and Schwille]{litschel2021reconstitution}
Litschel,~T.; Kelley,~C.~F.; Holz,~D.; Adeli~Koudehi,~M.; Vogel,~S.~K.;
  Burbaum,~L.; Mizuno,~N.; Vavylonis,~D.; Schwille,~P. Reconstitution of
  contractile actomyosin rings in vesicles. \emph{Nature communications}
  \textbf{2021}, \emph{12}, 2254\relax
\mciteBstWouldAddEndPuncttrue
\mciteSetBstMidEndSepPunct{\mcitedefaultmidpunct}
{\mcitedefaultendpunct}{\mcitedefaultseppunct}\relax
\EndOfBibitem
\bibitem[Bashirzadeh \latin{et~al.}(2021)Bashirzadeh, Redford, Lorpaiboon,
  Groaz, Moghimianavval, Litschel, Schwille, Hocky, Dinner, and
  Liu]{bashirzadeh2021actin}
Bashirzadeh,~Y.; Redford,~S.~A.; Lorpaiboon,~C.; Groaz,~A.; Moghimianavval,~H.;
  Litschel,~T.; Schwille,~P.; Hocky,~G.~M.; Dinner,~A.~R.; Liu,~A.~P. Actin
  crosslinker competition and sorting drive emergent {GUV} size-dependent actin
  network architecture. \emph{Communications Biology} \textbf{2021}, \emph{4},
  1136\relax
\mciteBstWouldAddEndPuncttrue
\mciteSetBstMidEndSepPunct{\mcitedefaultmidpunct}
{\mcitedefaultendpunct}{\mcitedefaultseppunct}\relax
\EndOfBibitem
\bibitem[Sakamoto and Murrell(2024)Sakamoto, and Murrell]{sakamoto2024f}
Sakamoto,~R.; Murrell,~M.~P. F-actin architecture determines the conversion of
  chemical energy into mechanical work. \emph{Nature Communications}
  \textbf{2024}, \emph{15}, 3444\relax
\mciteBstWouldAddEndPuncttrue
\mciteSetBstMidEndSepPunct{\mcitedefaultmidpunct}
{\mcitedefaultendpunct}{\mcitedefaultseppunct}\relax
\EndOfBibitem
\bibitem[Kurokawa \latin{et~al.}(2017)Kurokawa, Fujiwara, Morita, Kawamata,
  Kawagishi, Sakai, Murayama, Nomura, Murata, Takinoue, \latin{et~al.}
  others]{kurokawa2017dna}
Kurokawa,~C.; Fujiwara,~K.; Morita,~M.; Kawamata,~I.; Kawagishi,~Y.; Sakai,~A.;
  Murayama,~Y.; Nomura,~S.-i.~M.; Murata,~S.; Takinoue,~M., \latin{et~al.}
  {DNA} cytoskeleton for stabilizing artificial cells. \emph{Proceedings of the
  National Academy of Sciences} \textbf{2017}, \emph{114}, 7228--7233\relax
\mciteBstWouldAddEndPuncttrue
\mciteSetBstMidEndSepPunct{\mcitedefaultmidpunct}
{\mcitedefaultendpunct}{\mcitedefaultseppunct}\relax
\EndOfBibitem
\bibitem[Agarwal \latin{et~al.}(2021)Agarwal, Klocke, Pungchai, and
  Franco]{agarwal2021dynamic}
Agarwal,~S.; Klocke,~M.~A.; Pungchai,~P.~E.; Franco,~E. Dynamic self-assembly
  of compartmentalized DNA nanotubes. \emph{Nature communications}
  \textbf{2021}, \emph{12}, 3557\relax
\mciteBstWouldAddEndPuncttrue
\mciteSetBstMidEndSepPunct{\mcitedefaultmidpunct}
{\mcitedefaultendpunct}{\mcitedefaultseppunct}\relax
\EndOfBibitem
\bibitem[Sauter \latin{et~al.}(2023)Sauter, Schr{\"o}ter, Frey, Weber,
  Mersdorf, Janiesch, Platzman, and Spatz]{sauter2023artificial}
Sauter,~D.; Schr{\"o}ter,~M.; Frey,~C.; Weber,~C.; Mersdorf,~U.;
  Janiesch,~J.-W.; Platzman,~I.; Spatz,~J.~P. Artificial Cytoskeleton Assembly
  for Synthetic Cell Motility. \emph{Macromolecular Bioscience} \textbf{2023},
  \emph{23}, 2200437\relax
\mciteBstWouldAddEndPuncttrue
\mciteSetBstMidEndSepPunct{\mcitedefaultmidpunct}
{\mcitedefaultendpunct}{\mcitedefaultseppunct}\relax
\EndOfBibitem
\bibitem[Paulose \latin{et~al.}(2012)Paulose, Vliegenthart, Gompper, and
  Nelson]{paulose_fluctuating_2012}
Paulose,~J.; Vliegenthart,~G.~A.; Gompper,~G.; Nelson,~D.~R. Fluctuating shells
  under pressure. \emph{Proceedings of the National Academy of Sciences}
  \textbf{2012}, \emph{109}, 19551--19556, Publisher: Proceedings of the
  National Academy of Sciences\relax
\mciteBstWouldAddEndPuncttrue
\mciteSetBstMidEndSepPunct{\mcitedefaultmidpunct}
{\mcitedefaultendpunct}{\mcitedefaultseppunct}\relax
\EndOfBibitem
\bibitem[Maji and Rabin(2022)Maji, and Rabin]{maji_network_2022}
Maji,~A.; Rabin,~Y. Network model of active elastic shells swollen by
  hydrostatic pressure. \emph{Soft Matter} \textbf{2022}, \emph{18},
  7981--7989, Publisher: Royal Society of Chemistry\relax
\mciteBstWouldAddEndPuncttrue
\mciteSetBstMidEndSepPunct{\mcitedefaultmidpunct}
{\mcitedefaultendpunct}{\mcitedefaultseppunct}\relax
\EndOfBibitem
\bibitem[Ma and Berro(2019)Ma, and Berro]{ma_crosslinking_2019}
Ma,~R.; Berro,~J. Crosslinking actin networks produces compressive force.
  \emph{Cytoskeleton} \textbf{2019}, \emph{76}, 346--354\relax
\mciteBstWouldAddEndPuncttrue
\mciteSetBstMidEndSepPunct{\mcitedefaultmidpunct}
{\mcitedefaultendpunct}{\mcitedefaultseppunct}\relax
\EndOfBibitem
\bibitem[Akenuwa and Abel(2023)Akenuwa, and Abel]{akenuwa_organization_2023}
Akenuwa,~O.~H.; Abel,~S.~M. Organization and dynamics of cross-linked actin
  filaments in confined environments. \emph{Biophysical Journal} \textbf{2023},
  \emph{122}, 30--42\relax
\mciteBstWouldAddEndPuncttrue
\mciteSetBstMidEndSepPunct{\mcitedefaultmidpunct}
{\mcitedefaultendpunct}{\mcitedefaultseppunct}\relax
\EndOfBibitem
\bibitem[Adeli~Koudehi \latin{et~al.}(2019)Adeli~Koudehi, Rutkowski, and
  Vavylonis]{adeli_koudehi_organization_2019}
Adeli~Koudehi,~M.; Rutkowski,~D.~M.; Vavylonis,~D. Organization of associating
  or crosslinked actin filaments in confinement. \emph{Cytoskeleton}
  \textbf{2019}, \emph{76}, 532--548, \_eprint:
  https://onlinelibrary.wiley.com/doi/pdf/10.1002/cm.21565\relax
\mciteBstWouldAddEndPuncttrue
\mciteSetBstMidEndSepPunct{\mcitedefaultmidpunct}
{\mcitedefaultendpunct}{\mcitedefaultseppunct}\relax
\EndOfBibitem
\bibitem[Milchev \latin{et~al.}(2017)Milchev, Egorov, Nikoubashman, and
  Binder]{milchev_conformations_2017}
Milchev,~A.; Egorov,~S.~A.; Nikoubashman,~A.; Binder,~K. Conformations and
  orientational ordering of semiflexible polymers in spherical confinement.
  \emph{The Journal of Chemical Physics} \textbf{2017}, \emph{146},
  194907\relax
\mciteBstWouldAddEndPuncttrue
\mciteSetBstMidEndSepPunct{\mcitedefaultmidpunct}
{\mcitedefaultendpunct}{\mcitedefaultseppunct}\relax
\EndOfBibitem
\bibitem[Nikoubashman \latin{et~al.}(2017)Nikoubashman, Vega, Binder, and
  Milchev]{nikoubashman_semiflexible_2017}
Nikoubashman,~A.; Vega,~D.~A.; Binder,~K.; Milchev,~A. Semiflexible {Polymers}
  in {Spherical} {Confinement}: {Bipolar} {Orientational} {Order} {Versus}
  {Tennis} {Ball} {States}. \emph{Physical Review Letters} \textbf{2017},
  \emph{118}, 217803, Publisher: American Physical Society\relax
\mciteBstWouldAddEndPuncttrue
\mciteSetBstMidEndSepPunct{\mcitedefaultmidpunct}
{\mcitedefaultendpunct}{\mcitedefaultseppunct}\relax
\EndOfBibitem
\bibitem[Fo{\v{s}}nari{\v{c}} \latin{et~al.}(2013)Fo{\v{s}}nari{\v{c}},
  Igli{\v{c}}, Kroll, and May]{fovsnarivc2013monte}
Fo{\v{s}}nari{\v{c}},~M.; Igli{\v{c}},~A.; Kroll,~D.~M.; May,~S. Monte Carlo
  simulations of a polymer confined within a fluid vesicle. \emph{Soft Matter}
  \textbf{2013}, \emph{9}, 3976--3984\relax
\mciteBstWouldAddEndPuncttrue
\mciteSetBstMidEndSepPunct{\mcitedefaultmidpunct}
{\mcitedefaultendpunct}{\mcitedefaultseppunct}\relax
\EndOfBibitem
\bibitem[Kierfeld \latin{et~al.}(2010)Kierfeld, Baczynski, Gutjahr, Kühne, and
  Lipowsky]{kierfeld_modelling_2010}
Kierfeld,~J.; Baczynski,~K.; Gutjahr,~P.; Kühne,~T.; Lipowsky,~R. Modelling
  semiflexible polymers : shape analysis, buckling instabilities, and force
  generation. \emph{Soft Matter} \textbf{2010}, \emph{6}, 5764--5769,
  Publisher: Royal Society of Chemistry\relax
\mciteBstWouldAddEndPuncttrue
\mciteSetBstMidEndSepPunct{\mcitedefaultmidpunct}
{\mcitedefaultendpunct}{\mcitedefaultseppunct}\relax
\EndOfBibitem
\bibitem[Claessens \latin{et~al.}(2006)Claessens, Tharmann, Kroy, and
  Bausch]{claessens_microstructure_2006}
Claessens,~M. M. a.~E.; Tharmann,~R.; Kroy,~K.; Bausch,~A.~R. Microstructure
  and viscoelasticity of confined semiflexible polymer networks. \emph{Nature
  Physics} \textbf{2006}, \emph{2}, 186--189, Publisher: Nature Publishing
  Group\relax
\mciteBstWouldAddEndPuncttrue
\mciteSetBstMidEndSepPunct{\mcitedefaultmidpunct}
{\mcitedefaultendpunct}{\mcitedefaultseppunct}\relax
\EndOfBibitem
\bibitem[Shi \latin{et~al.}(2023)Shi, Zou, Wu, Wang, Zhang, Gao, and
  Yi]{shi2023morphological}
Shi,~C.; Zou,~G.; Wu,~Z.; Wang,~M.; Zhang,~X.; Gao,~H.; Yi,~X. Morphological
  transformations of vesicles with confined flexible filaments.
  \emph{Proceedings of the National Academy of Sciences} \textbf{2023},
  \emph{120}, e2300380120\relax
\mciteBstWouldAddEndPuncttrue
\mciteSetBstMidEndSepPunct{\mcitedefaultmidpunct}
{\mcitedefaultendpunct}{\mcitedefaultseppunct}\relax
\EndOfBibitem
\bibitem[Vliegenthart and Gompper(2011)Vliegenthart, and
  Gompper]{vliegenthart_compression_2011}
Vliegenthart,~G.~A.; Gompper,~G. Compression, crumpling and collapse of
  spherical shells and capsules. \emph{New Journal of Physics} \textbf{2011},
  \emph{13}, 045020\relax
\mciteBstWouldAddEndPuncttrue
\mciteSetBstMidEndSepPunct{\mcitedefaultmidpunct}
{\mcitedefaultendpunct}{\mcitedefaultseppunct}\relax
\EndOfBibitem
\bibitem[Šarić and Cacciuto(2013)Šarić, and
  Cacciuto]{saric_self-assembly_2013}
Šarić,~A.; Cacciuto,~A. Self-assembly of nanoparticles adsorbed on fluid and
  elastic membranes. \emph{Soft Matter} \textbf{2013}, \emph{9}, 6677--6695,
  Publisher: The Royal Society of Chemistry\relax
\mciteBstWouldAddEndPuncttrue
\mciteSetBstMidEndSepPunct{\mcitedefaultmidpunct}
{\mcitedefaultendpunct}{\mcitedefaultseppunct}\relax
\EndOfBibitem
\bibitem[Weeks \latin{et~al.}(1971)Weeks, Chandler, and
  Andersen]{weeks_role_1971}
Weeks,~J.~D.; Chandler,~D.; Andersen,~H.~C. Role of {Repulsive} {Forces} in
  {Determining} the {Equilibrium} {Structure} of {Simple} {Liquids}. \emph{The
  Journal of Chemical Physics} \textbf{1971}, \emph{54}, 5237--5247\relax
\mciteBstWouldAddEndPuncttrue
\mciteSetBstMidEndSepPunct{\mcitedefaultmidpunct}
{\mcitedefaultendpunct}{\mcitedefaultseppunct}\relax
\EndOfBibitem
\bibitem[Zhang \latin{et~al.}(2019)Zhang, Peng, Liu, Jiang, Ji, and
  Shen]{zhang_persistence_2019}
Zhang,~J.-Z.; Peng,~X.-Y.; Liu,~S.; Jiang,~B.-P.; Ji,~S.-C.; Shen,~X.-C. The
  {Persistence} {Length} of {Semiflexible} {Polymers} in {Lattice} {Monte}
  {Carlo} {Simulations}. \emph{Polymers} \textbf{2019}, \emph{11}, 295, Number:
  2 Publisher: Multidisciplinary Digital Publishing Institute\relax
\mciteBstWouldAddEndPuncttrue
\mciteSetBstMidEndSepPunct{\mcitedefaultmidpunct}
{\mcitedefaultendpunct}{\mcitedefaultseppunct}\relax
\EndOfBibitem
\bibitem[Rovigatti \latin{et~al.}(2018)Rovigatti, Russo, and
  Romano]{rovigatti_how_2018}
Rovigatti,~L.; Russo,~J.; Romano,~F. How to simulate patchy particles.
  \emph{The European Physical Journal E} \textbf{2018}, \emph{41}, 1--12\relax
\mciteBstWouldAddEndPuncttrue
\mciteSetBstMidEndSepPunct{\mcitedefaultmidpunct}
{\mcitedefaultendpunct}{\mcitedefaultseppunct}\relax
\EndOfBibitem
\bibitem[Finkelstein \latin{et~al.}(2020)Finkelstein, Fiorin, and
  Seibold]{finkelstein_comparison_2020}
Finkelstein,~J.; Fiorin,~G.; Seibold,~B. Comparison of modern {Langevin}
  integrators for simulations of coarse-grained polymer melts. \emph{Molecular
  Physics} \textbf{2020}, \emph{118}, e1649493, Publisher: Taylor \& Francis
  \_eprint: https://doi.org/10.1080/00268976.2019.1649493\relax
\mciteBstWouldAddEndPuncttrue
\mciteSetBstMidEndSepPunct{\mcitedefaultmidpunct}
{\mcitedefaultendpunct}{\mcitedefaultseppunct}\relax
\EndOfBibitem
\bibitem[Li \latin{et~al.}(2018)Li, Zhu, Lu, and Sun]{li_general_2018}
Li,~Z.~W.; Zhu,~Y.~L.; Lu,~Z.~Y.; Sun,~Z.~Y. General patchy ellipsoidal
  particle model for the aggregation behaviors of shape- and/or
  surface-anisotropic building blocks. \emph{Soft Matter} \textbf{2018},
  \emph{14}, 7625--7633, Publisher: Royal Society of Chemistry\relax
\mciteBstWouldAddEndPuncttrue
\mciteSetBstMidEndSepPunct{\mcitedefaultmidpunct}
{\mcitedefaultendpunct}{\mcitedefaultseppunct}\relax
\EndOfBibitem
\bibitem[Limozin and Sackmann(2002)Limozin, and
  Sackmann]{limozin2002polymorphism}
Limozin,~L.; Sackmann,~E. Polymorphism of cross-linked actin networks in giant
  vesicles. \emph{Physical Review Letters} \textbf{2002}, \emph{89},
  168103\relax
\mciteBstWouldAddEndPuncttrue
\mciteSetBstMidEndSepPunct{\mcitedefaultmidpunct}
{\mcitedefaultendpunct}{\mcitedefaultseppunct}\relax
\EndOfBibitem
\bibitem[Baumann and Surrey(2014)Baumann, and Surrey]{baumann2014motor}
Baumann,~H.; Surrey,~T. Motor-mediated cortical versus astral microtubule
  organization in lipid-monolayered droplets. \emph{Journal of Biological
  Chemistry} \textbf{2014}, \emph{289}, 22524--22535\relax
\mciteBstWouldAddEndPuncttrue
\mciteSetBstMidEndSepPunct{\mcitedefaultmidpunct}
{\mcitedefaultendpunct}{\mcitedefaultseppunct}\relax
\EndOfBibitem
\bibitem[Theriot(2000)]{theriot_polymerization_2000}
Theriot,~J.~A. The {Polymerization} {Motor}. \emph{Traffic} \textbf{2000},
  \emph{1}, 19--28\relax
\mciteBstWouldAddEndPuncttrue
\mciteSetBstMidEndSepPunct{\mcitedefaultmidpunct}
{\mcitedefaultendpunct}{\mcitedefaultseppunct}\relax
\EndOfBibitem
\bibitem[Peskin \latin{et~al.}(1993)Peskin, Odell, and
  Oster]{peskin_cellular_1993}
Peskin,~C.~S.; Odell,~G.~M.; Oster,~G.~F. Cellular motions and thermal
  fluctuations: the {Brownian} ratchet. \emph{Biophysical Journal}
  \textbf{1993}, \emph{65}, 316--324\relax
\mciteBstWouldAddEndPuncttrue
\mciteSetBstMidEndSepPunct{\mcitedefaultmidpunct}
{\mcitedefaultendpunct}{\mcitedefaultseppunct}\relax
\EndOfBibitem
\bibitem[Mogilner and Oster(1996)Mogilner, and Oster]{mogilner_cell_1996}
Mogilner,~A.; Oster,~G. Cell motility driven by actin polymerization.
  \emph{Biophysical Journal} \textbf{1996}, \emph{71}, 3030--3045\relax
\mciteBstWouldAddEndPuncttrue
\mciteSetBstMidEndSepPunct{\mcitedefaultmidpunct}
{\mcitedefaultendpunct}{\mcitedefaultseppunct}\relax
\EndOfBibitem
\bibitem[Dmitrieff and Nédélec(2016)Dmitrieff, and
  Nédélec]{dmitrieff_amplification_2016}
Dmitrieff,~S.; Nédélec,~F. Amplification of actin polymerization forces.
  \emph{The Journal of Cell Biology} \textbf{2016}, \emph{212}, 763--766\relax
\mciteBstWouldAddEndPuncttrue
\mciteSetBstMidEndSepPunct{\mcitedefaultmidpunct}
{\mcitedefaultendpunct}{\mcitedefaultseppunct}\relax
\EndOfBibitem
\end{mcitethebibliography}

\clearpage

\appendix
\section{Supplemental Information to generating forces in confinement via polymerization}

\subsection{Deformations of empty shells}
\begin{figure}[!h]
    \begin{center}
    \includegraphics[width=180mm]{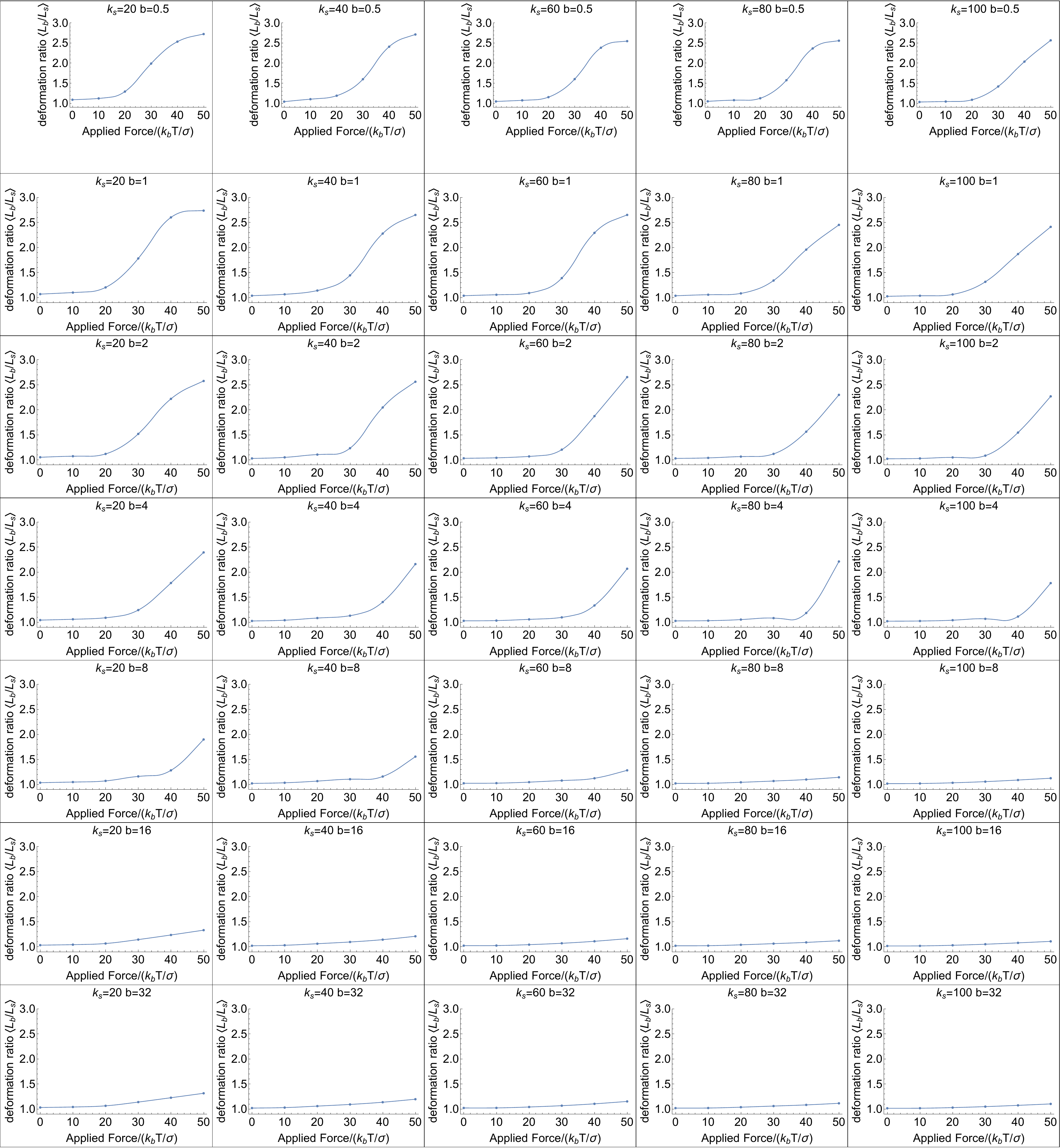}
    \caption{ Response of the elastic shell to applied force on two monomers on opposite ends of the shell. We apply a force of the given magnitude to each monomer, pushing in opposite directions. We measure the axial ratios under different strengths of the pulling force, across different elastic parameters for the shell. Increasing the stretching and bending moduli of the shell increases its resistence to pulling (higher effective spring constants)  }
    \label{fig:sfig1} %
    \end{center}
    \end{figure}
    \clearpage

\subsection{Bundling transitions in toy models}

We plot the bundling transition in the toy model for different values of the shell bending potential.

\begin{figure}[!h]
    \begin{center}
    \includegraphics[width=90mm]{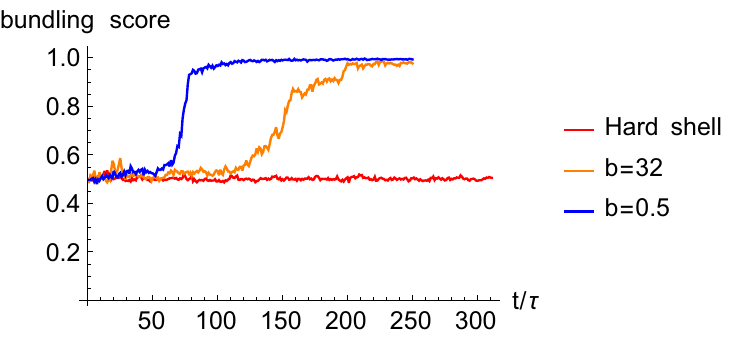}
    \caption{ The bundling parameter (defined in the main text) for the toy model of polymers growing in shells with $k_s=20$ and different bending moduli, or in a completely hard shell. In the hard shell, there is no bundling transition.    }
    \label{fig:sfig2} %
    \end{center}
    \end{figure}

\clearpage
\subsection{Persistence length of free polymers}

The main text used polymers with different microscopic parameters for the binding energy and binding angles of our monomeric subunits. Below, we present what the persistence lengths of polymers in free space are for different choices of microscopic parameters $\delta E$ and $\theta_m$ (the binding energy and angular aperture of the binding site, respectively). The persistence length is calculated as the correlation distance of tangent vectors to the polymer:

\begin{equation}
\langle \cos(\theta )\rangle = e^{-L/P}
\end{equation}
where $L$ is the distance along the contour of the chain and P is the persistence length.

We perform these simulations by beginning with all the monomers in free space and letting them grow naturally, then measuring the average persistence length of the resultant polymers after equilibration. The persistence length is presented in the following figure:

\begin{figure}[!h]
    \begin{center}
    \includegraphics[width=90mm]{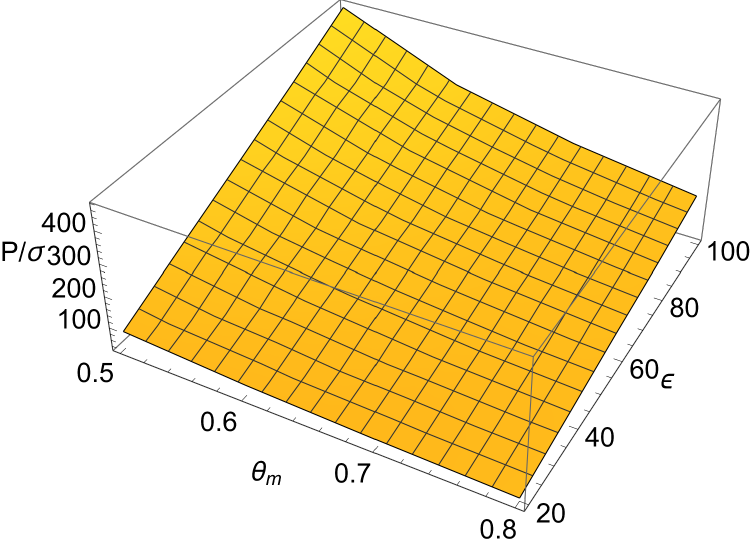}
    \caption{ The persistence lengths of polymers with different microscopic parameters, measured in terms of the hard sphere diameter. The persistence length increases with increasing binding energy, and decreases with greater patch size.   }
    \label{fig:sfig3} %
    \end{center}
    \end{figure}

as can be observed, the persistence lengths of all the polymers in free space are larger than the sizes of the vesicles we are embedding them in, leading to the polymers exerting elastic forces on the vesicle. The presence of the vesicle will affect measured persistence lengths, by forcing the polymers to bend, but at an energetic penalty.

\clearpage
\subsection{Phase diagrams of self assembly polymer models against microscopic shell and monomer parameters}

\begin{figure}[!h]
    \begin{center}
    \includegraphics[width=90mm]{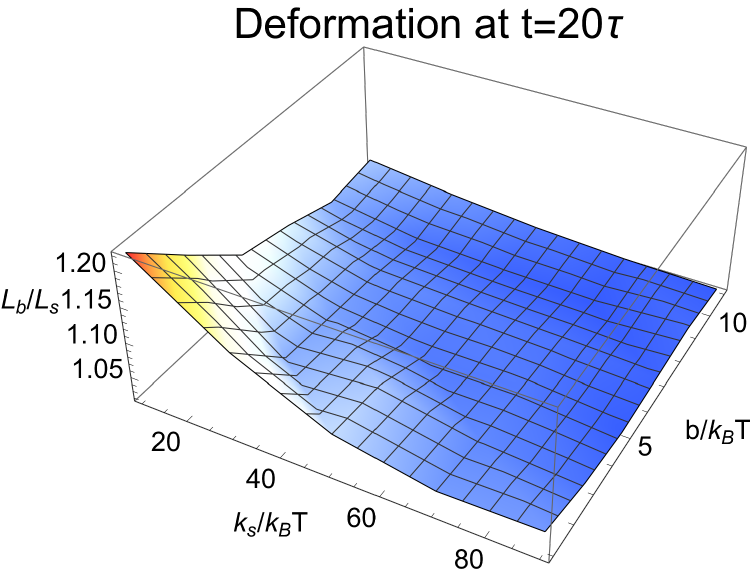}
    \caption{ The deformation score for the self assembling polymer system at the given time when the shell parameters are varied for $\epsilon=100,\theta_m=0.6$. As with the toy model, the weaker shells have more axial deformation.    }
    \label{fig:sfig4} %
    \end{center}
    \end{figure}

\begin{figure}[!h]
    \begin{center}
    \includegraphics[width=90mm]{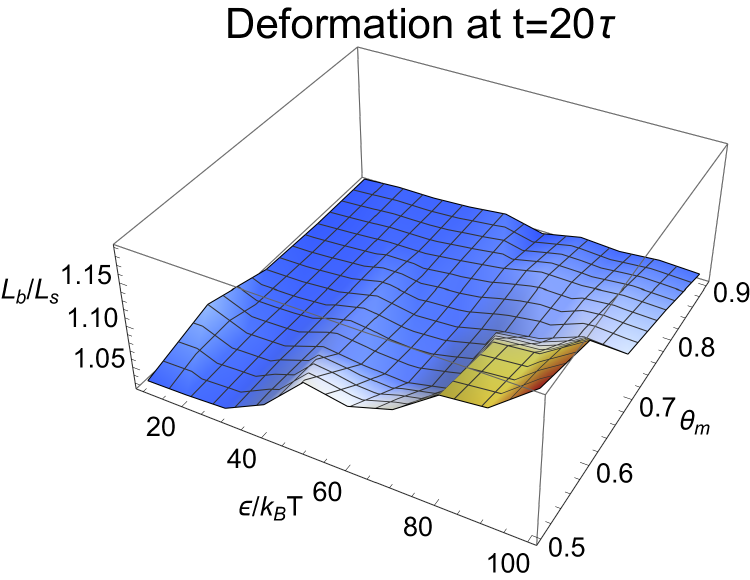}
    \caption{  The deformation score at a given time when the monomer parameters are varied. For $k_s=20,b=4,R\tau=50$. Stiffer polymers always do better, despite being more difficult to assemble.     }
    \label{fig:sfig5} %
    \end{center}
    \end{figure}

\clearpage
\subsection{Additional results for different multivalent schemes}
\begin{figure}[!h]
    \begin{center}
    \includegraphics[width=90mm]{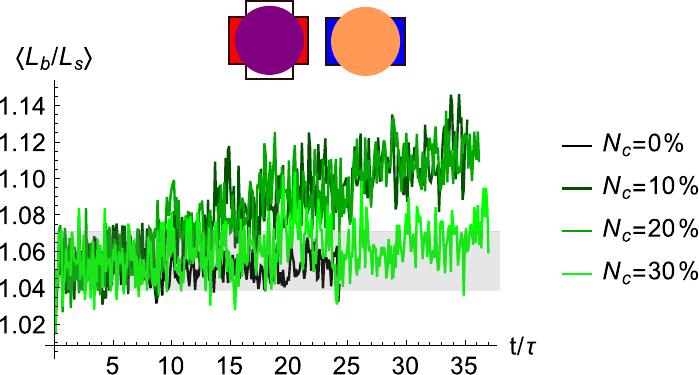}
    \caption{  The average deformation scores against monomer numbers as the total number of crosslinkers are varied for a variant of main text scheme 4D) where each monomer has two potential crosslinking sites on the surface as indicated. The results are not qualitatively different to the main text scheme.  $k_s=20,b=1,R\tau=50,\epsilon=50,\theta_m=0.8$     }
    \label{fig:sfig6} %
    \end{center}
    \end{figure}

    \begin{figure}[!h]
    \begin{center}
    \includegraphics[width=90mm]{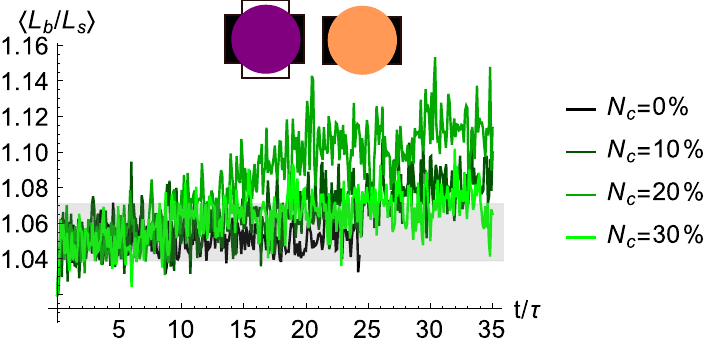}
    \caption{ The average deformation scores against monomer numbers as the total number of crosslinkers are varied for a variant of main text scheme 4D) where crosslinkers have the ability to bind to one another. One can still observe deformation, but the relative proportion of crosslinkers to achieve optimal performance is modified.  $k_s=20,b=1,R\tau=50,\epsilon=50,\theta_m=0.8$    }
    \label{fig:sfig7} %
    \end{center}
    \end{figure}

\clearpage
\subsection{Assembly under steric hindrance}

In the main text, we introduced a model of assembly where monomers are added to the interior of the shell. One could imagine an alternative scheme where all the monomers are already present in the shell and are ``activated" (i.e., the binding sites are turned on) at some rate $R$. While we do not expect this model to be too different to our main model, potential steric interactions with unactivated monomers could lead to potential differences in results. We show here that this model variant still produces axial deformation. 

    \begin{figure}[!h]
    \begin{center}
    \includegraphics[width=90mm]{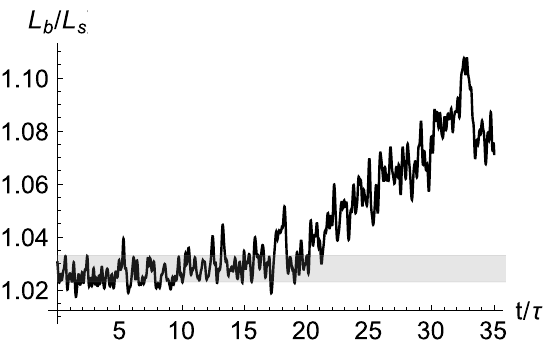}
    \caption{ Model variant where the monomers are already present in the shell, but need to be activated at some rate $R$. We still observe axial deformation in this case, the steric hindrance of the already present monomers does not modify the ability of the polymerization mechanism to generate force.  $k_s=20,b=4,R\tau=50,\epsilon=100,\theta_m=0.6$  }
    \label{fig:sfig8} %
    \end{center}
    \end{figure}

\end{document}